\documentclass[twocolumn]{aastex631}

\graphicspath{{./}{figures/}}

\begin{document}

\title{Coronal Magnetic Field Extrapolation and Topological Analysis of Fine-Scale Structures during Solar Flare Precursors}

\correspondingauthor{Qiang Hu}
\email{qh0001@uah.edu}

\author[0000-0001-8749-1022]{Wen He}
\affiliation{Department of Space Science,\\
The University of Alabama in Huntsville, Huntsville, AL 35805, USA}

\author[0000-0002-7570-2301]{Qiang Hu}
\affiliation{Department of Space Science,\\
The University of Alabama in Huntsville, Huntsville, AL 35805, USA}
\affiliation{Center for Space Plasma and Aeronomic Research (CSPAR), \\
The University of Alabama in Huntsville, Huntsville, AL 35805, USA}

\author[0000-0002-8179-3625]{Ju Jing}  
\affiliation{Center for Solar-Terrestrial Research,\\
New Jersey Institute of Technology,
University Heights, Newark, NJ 07102-1982, USA}

\author[0000-0002-5233-565X]{Haimin Wang}  
\affiliation{Center for Solar-Terrestrial Research,\\
New Jersey Institute of Technology, 
University Heights, Newark, NJ 07102-1982, USA}

\author[0000-0002-7018-6862]{Chaowei Jiang}
\affiliation{Institute of Space Science and Applied Technology, \\
Harbin Institute of Technology, Shenzhen, China }

\author[0000-0002-4241-627X]{Sushree S. Nayak}
 \affiliation{Center for Space Plasma and Aeronomic Research (CSPAR), \\
 The University of Alabama in Huntsville, Huntsville, AL 35805, USA}

\author[0000-0003-0819-464X]{Avijeet Prasad}
\affiliation{Institute of Theoretical Astrophysics, \\
University of Oslo, Postboks 1029 Blindern, 0315 Oslo, Norway}
\affiliation{Rosseland Centre for Solar Physics, \\
University of Oslo, Postboks 1029 Blindern, 0315 Oslo, Norway}

\begin{abstract}
Magnetic field plays an important role in various solar eruptions like flares, coronal mass ejections, etc. The formation and evolution of characteristic magnetic field topology in solar eruptions are critical problems that will ultimately help us understand the origination of these eruptions in the solar source regions. With the development of advanced techniques and instruments, observations with higher resolutions in different wavelengths and fields of view have provided more quantitative information for finer structures. So it is essential to improve our method to study the magnetic field topology in the solar source regions by taking advantage of high-resolution observations. In this study, we employ a  nonlinear force-free field (NLFFF) extrapolation method based on a nonuniform grid setting for an M-class flare eruption event (SOL2015-06-22T17:39) with embedded magnetograms from the Solar Dynamics Observatory (SDO) and the Goode Solar Telescope (GST). The extrapolation results employing the embedded magnetogram for the bottom boundary are obtained by maintaining the native resolutions of the corresponding GST and SDO magnetograms. We compare the field line connectivity with the simultaneous GST/H$\alpha$ and SDO/AIA observations for fine-scale structures associated with precursor brightenings. Then we perform a topological analysis of the field line connectivity corresponding to fine-scale magnetic field structures based on the extrapolation results. The results indicate that by combining the high-resolution GST magnetogram with a larger HMI magnetogram, the derived magnetic field topology is consistent with a scenario of magnetic reconnection among sheared field lines across the main polarity inversion line during solar flare precursors. 
\end{abstract}



\section{Introduction}
There are different kinds of spectacular eruptions in the solar atmosphere, such as flares, coronal mass ejections (CMEs) and jets, which release energy in various spatial and temporal scales. In particular, solar flare eruption attracts a lot of attention among these eruption phenomena as an explosive, energetic phenomenon with enhanced emission throughout the electromagnetic spectrum in a dynamic and complicated process. In multi-wavelength observations, a flare usually goes through three major phases, namely the preflare, the impulsive and the gradual phases. And the life of a flare spans from tens of seconds to several hours \citep[see a review by][]{2017Benz}. During the flare eruption, the energy release could be as large as $10^{32}$ ergs, while the major contribution comes from the magnetic energy comparing to other sources. To figure out the underlying physical mechanisms, i.e., the source of energy for release, a lot of efforts have been made on the different perspectives of the main phase (impulsive and gradual phases) of solar flares. For example, the standard two-dimensional (2D) flare model  \citep[so-called CSHKP model,][]{1964Carmichael, 1966Sturrock, 1974Hirayama, 1976KoppPneuman} proposes that magnetic reconnection plays a major role in the energy release during the evolution of a flare. 

In addition to the main phase of the flare eruption, it is noteworthy that there are also some interesting small-scale localized energy release phenomena in the precursor phase (before the flare main phase or before the time of the peak X-ray flux emission), e.g., the so-called flare precursor brightenings. \citet{1959Bumba} introduced the concept of flare precursors which were observed as a short-term and small brightening before the main flare onset. Later it has been observed in many flares through multi wavelengths including X-ray, optical, ultraviolet/extreme ultraviolet (UV/EUV) and microwave observations \citep{2014Awasthi, 2014Bamba, 2017Bamba}. \citet{1991Tappin} performed a statistical study based on X-ray observations to investigate the correlation between flare precursors and flare onsets and they summarized that most flares as measured by X-ray emissions are preceded by one or more soft X-ray precursors with 10 to 60 minutes prior to the flare onset \citep[see also a more recent statistical study by][]{2016Gyenge}. Later on, \citet{2007Chifor} reported that the precursors locate near or on the polarity inversion line (PIL) and hard X-ray precursor brightenings move rapidly along a PIL before the flare main phase based on the analysis of a list of preflare events by combining multi-wavelength observations with the evolution of photospheric magnetic fields. Their study also provides evidence of the spatial and temporal correlation between the preflare activities and the filament eruption onsets. Therefore, the investigation of the flare precursors is an important subject not only for the initiation mechanism of flares but also for the associated eruption phenomena. Given the relatively smaller-scale energy release of flare precursors with respect to the main phase of the flare evolution, observations at higher spatial and temporal resolutions are required. In the meantime, it is also essential to validate different eruption initiation mechanisms with a better understanding of the magnetic field topology for flare precursors. However, the inherent fine-scale three-dimensional (3D) magnetic field topology change is still unclear owing to a lack of the quantitative study by using high-resolution vector magnetograms. Here we intend to perform an analysis of the fine-scale magnetic field structures associated with flare precursor brightenings through nonlinear force-free field (NLFFF) extrapolations by employing recently available high-resolution vector magnetograms from multiple sources.

With the recent development of observational techniques and instruments, more advanced high-resolution solar observational data become available, including space-based telescopes like the Solar Dynamics Observatory (SDO), the Hinode satellite, and the Solar Orbiter, and also the ground-based telescopes like the 1.6-meter Goode Solar Telescope (GST), the 4-meter Daniel K. Inouye Solar Telescope (DKIST) and so on. More advanced observations will be definitely crucial to improving our understanding of small-scale energy release process like flare precursors and their connections to the following flare main phase. Therefore, there is a growing demand for taking full advantage of data from multiple instruments with necessary improvements to the existing methods. For instance, the ultra high-resolution observation in a smaller field of view (FOV) can contribute to the analysis of the fine-scale structure of small-scale events \citep{2016Jing, 2017Wang, 2022Zhao}. Alternatively, a relatively lower-resolution observation in a larger FOV has more advantage to extend the spatial coverage and to describe the magnetic connections to surrounding structures. As for the analysis of coronal magnetic structures in solar eruptions, vector magnetograms may be obtained from multiple instruments with different spatio-temporal resolutions and FOVs (mostly on the photosphere). One desirable approach is to be able to combine these vector magnetograms for the numerical extrapolation of the coronal magnetic field, while preserving their respective advantages.

Numerical simulations can be a viable tool to derive unavailable data like the 3D coronal magnetic fields with reasonable assumptions \citep{2022Jiang}. For example, the photospheric vector magnetograms are often employed as bottom boundary conditions (BCs) in different numerical simulation methods to reconstruct the 3D coronal magnetic field in the solar source region. However, the limitations of computational resources for the numerical simulation capability of a more realistic full magnetohydrodynamics (MHD) model also lead to the use of the NLFFF extrapolation method to reconstruct static 3D coronal magnetic field based on a force-free assumption \citep{2021Wiegelmann, 2012Jiang}. Different kinds of numerical methods have been proposed to reconstruct the NLFFF for the coronal magnetic field from specific BCs and sometimes pseudo-initial conditions, including the upward integration, Grad-Rubin iteration, MHD relaxation, optimization approach, and so on \citep[see a review by][]{2021Wiegelmann}. The computation speed and quality of different numerical modeling results may vary significantly for the realistic solar magnetograms not only due to the differences in algorithms and their specific realizations, but also the quality of the input magnetograms. For instance, the spatial resolution of input vector magnetograms as the bottom BC has been proven to have an important effect on NLFFF extrapolation results, including the magnetic energy and associated magnetic field topology, as reported by \citet{2013Thalmann} and \citet{2015DeRosa}. In the meanwhile, more solar observations become available and can provide vector magnetograms in different spatial resolutions and FOVs for the same solar source region. Therefore, in order to improve the computational efficiency and take full advantage of available observations, it becomes natural to incorporate the available higher-resolution magnetograms (often in a small FOV) into the bottom BC along with the lower-resolution magnetograms (with a larger FOV) to conduct the extrapolation, thus maintaining the native resolution of the higher-resolution magnetogram and a larger FOV at the same time, especially for the study of fine-scale structure in flare precursors. 

An M6.5 class flare erupted close to the solar disk center ($8 \degr$W 12$\degr$N) on June 22, 2015 in active region NOAA 12371. The impulsive phase of the flare starts at $\sim$ 17:51 UT. Two short-duration small-scale brightenings were observed in unprecedented spatio-temporal resolution by the 1.6-m GST along with photospheric magnetic field dynamics and reported as flare precursors by \citet{2017Wang}. That study focused on the two short episodes of flare precursors by utilizing high-resolution H$\alpha$ and photospheric magnetic field from GST observations, complemented by X-ray and microwave observations. And it indicates the evidence of successive reconnection process during the evolution of the precursor periods, which may contribute to the onset of the main flare. Many studies have been performed for this event in terms of different physical processes. The fine-scale structure of this flare and associated large-scale dynamic motion of flare ribbons have been discussed by \citet{2016Jing, 2017Jing}. \citet{2018Liu} and \citet{2018Xu} looked into the relationship between the flaring signatures and evolution of photospheric vector magnetic fields by taking advantage of the GST observations. For the flare onset process, some authors \citep{2018Awasthi, 2019Kang, 2022LiuYanjie} have studied the pre-eruptive magnetic configuration with reconstructed 3D magnetic field by the NLFFF extrapolation method based on the SDO/Helioseismic and Magnetic Imager \citep[HMI, ][]{2012Schou} magnetograms. In addition, a multi-instrument comparative study was also conducted by \citet{2022LiuNian}, which offers a quantitative description of the thermal behaviors for flare precursor over a large temperature range. For all these studies, it is always essential to compare the derived magnetic field configuration with the corresponding multi-wavelength imaging observations to help validate and interpret the extrapolation results when applicable.

In this study, we apply a type of MHD relaxation method with a conservation-element/solution-element (CESE) solver, so called the CESE-MHD-NLFFF method \citep[see details in][]{2011Jiang, 2013Jiang}, to obtain the 3D coronal magnetic field in an approximate force-free state. It has been applied to the analysis of magnetic field topology with realistic solar magnetic field data \citep{2013Jiang, 2017Duan, 2019Duan, 2022He}. For example, in our previous study, this method was applied to characterize the properties of magnetic flux ropes (MFRs) on the Sun and the results are then compared to the properties of the corresponding interplanetary counterparts quantitatively \citep{2022He}. The results indicated the importance of flare associated magnetic reconnection process in that the magnetic reconnection flux estimated from the analysis corresponds well to the magnetic flux content found in the MFR formed during the main phase of solar flares. A subsequent study \citep{hu2022validation} further implied the variability in the magnetic field topology changes of an MFR as manifested in the analysis of multiple observations of the associated flare/CME eruption process. For the present study, with the available high-resolution GST observations for the aforementioned M6.5 flare, we develop a modified version of the existing CESE-MHD-NLFFF code for embedded magnetograms by incorporating the high-resolution GST magnetogram and the larger-FOV SDO/HMI magnetogram as the bottom boundary condition with non-uniform grid spacing. The results will be compared to the extrapolations with single-set uniform magnetograms and the associated observations, mainly the high-resolution GST/H$\alpha$ images during the flare precursors. 

The article is organized as follows. First, the instrumentation and data used in this paper are described in Section \ref{data}. Then we present the modified CESE-MHD-NLFFF method and the associated convergence study in Section \ref{sec_method}. In Section \ref{sec_results}, the magnetic field topology from the extrapolations with different bottom BCs are presented and investigated in detail. The major results are summarized and conclusions are given in Section \ref{sec_summary}. 

\section{Instrumentation and Data} \label{data}
For this event, we make use of the observational data from both SDO and GST. SDO can provide full disk observations of the Sun routinely. Specifically, the Space-weather HMI Active Region Patch \citep[SHARP;][]{2014Bobra} vector magnetograms are used as the input bottom BCs of the NLFFF extrapolations. The SHARP data product offers photospheric vector magnetograms with a spatial resolution of $0.5''$/pixel ($\sim$ 365 km) in a cadence of 720 seconds. On the other hand, the corresponding remote sensing observations in UV and EUV wavelength channels are provided by the Atmospheric Imaging Assembly \citep[AIA,][]{2012Lemen} onboard SDO with a spatial resolution of around 0.6$''$/pixel ($\sim$ 438 km) and a moderate time cadence (12 s for EUV channels and 24 s for UV channels). For this 22 June 2015 flare event, the GST at the Big Bear Solar Observatory (BBSO) also obtained high-resolution observations during $\sim$ 16:50 -- 23:00 UT \citep{2016Jing, 2017Wang}. The $H\alpha$ images at the line center and off-bands ($\pm0.6$\r{A} and $\pm1.0$ \r{A}) are taken by the Visible Imaging Spectrometer \citep[VIS;][]{2010Cao} with a FOV of $\sim 57'' \times 64''$. The GST/VIS observations have a spatial resolution at $\sim 0.03''$/pixel (around 20km) and a time cadence of 28 s. The Near InfraRed Imaging Spectropolarimeter \citep[NIRIS;][]{2012Cao} of GST, equipped with the infrared detector and the dual Fabry-Perot interferometers system, provides the spectropolarimetric data (at Fe~{\sc i} 1565~nm doublet, 0.2~\AA\ bandpass). The spectropolarimetric data are processed with the NIRIS data processing pipeline including dark and flat field corrections, instrument crosstalk calibration and Milne-Eddington Stokes inversion, from which the vector magnetic fields could be extracted. The pixel size and temporal cadence of the resulting vector magnetograms are ~0.08$''$/pixel and 87 s, respectively. In general, the space-borne instrumentation provides a larger FOV and more continuous observations while the ground-based counterpart has a relatively smaller FOV and more sporadic temporal coverage, but has a much higher spatial resolution. Therefore it is desirable to combine these two sets of observations in order to make the best use of their data products, by embedding the higher resolution GST/NIRIS magnetogram into the corresponding part of the lower resolution, larger FOV SDO/HMI magnetogram.

To generate such a comprehensive NIRIS-HMI magnetogram, the first and most important step is the time-consuming alignment between the NIRIS and HMI magnetograms. Using the data here as an example, the FOV of the NIRIS magnetogram is $52'' \times 52''$, consisting of 650$\times$650 pixels with a spatial sampling of 0.08$''$/pixel, while the FOV of HMI is about $200'' \times 200''$, consisting of 400$\times$400 pixels, $0.5''$/pixel. By manual alignment, we find the exact position, with subpixel precision, of the NIRIS magnetogram of a small FOV on the HMI magnetogram of a large FOV. Since the spatial sampling rates of the two magnetograms are different, we interpolate the size of the HMI magnetogram to 2500$\times$2500 pixels with the same FOV, so that each of its pixels has the same spatial scale as the NIRIS magnetogram, i.e., 0.08$''$/pixel. Then, we embed the GST/NIRIS magnetogram in the middle of such an HMI magnetogram. After these steps, an embedded magnetogram with a FOV of $200''\times 200''$ and a pixel size of 0.08$''$/pixel is derived containing data from both HMI and NIRIS. We will construct a nonuniform grid structure for the NLFFF extrapolations and utilize the embedded magnetograms as our bottom BCs. 

\section{NLFFF Extrapolation Code for Embedded Magnetograms} \label{sec_method}

\subsection{Extrapolation Method by the CESE-MHD-NLFFF Code}
The CESE-MHD-NLFFF code is similar to a magnetofrictional method, which can be regarded as a special case of the MHD relaxation method. And it's mainly designed to solve the modified momentum equation and the magnetic induction equation \citep{2012Jiang, 2013Jiang},
\begin{equation}\label{eq1}
  \frac{\partial (\rho \textbf{v})}{\partial t} =(\nabla \times \textbf{B})\times \textbf{B} -\nu \rho \textbf{v},\quad \rho =|\textbf{B}|^{2} +\rho_{0},
\end{equation}
\begin{equation}\label{eq2}
  \frac{\partial \textbf{B}}{\partial t} = \nabla \times(\textbf{v} \times \textbf{B}) 
  -\textbf{v} \nabla \cdot \textbf{B} + \nabla(\mu \nabla \cdot \textbf{B}).
\end{equation}
These equations are solved as a kind of Dirichlet type boundary value problem. Based on the force-free field assumption (plasma $\beta \ll 1$), the magnetic force plays a major role. So other forces including the plasma pressure, the gravity and inertial forces can be ignored. In order to balance the Lorentz force, an artificial term $\nu \rho \textbf{v}$ in a frictional force form involving velocity \textbf{v} is added in the momentum equation (\ref{eq1}). In addition, a pseudo mass density $\rho$ is assumed to take the form given. And $\rho$ is modified with a small value $\rho_{0}$, e.g., $\rho_{0}=0.1$ (in the same unit as $|\textbf{B}|^{2}$), to deal with the case of very weak magnetic field. For the magnetic induction equation, two extra terms are added to control the divergence of the magnetic field. The equations (\ref{eq1}) and (\ref{eq2}) are solved through the iteration process until a converged solution of a quasi-static equilibrium state is approached.

The computation proceeds by iterations until a converged solution is reached as judged by a series of metrics. For the NLFFF extrapolation, a well-known problem is that the force-free condition may not always be satisfied in the inhomogeneous solar atmosphere, especially on the photosphere \citep{ 2001Gary}. \citet{2006Wiegelmann} proposed that a more consistent bottom BC for an NLFFF extrapolation can be obtained by modifying the original photospheric magnetogram to mimic a force-free chromospheric magnetogram. Such a practice commonly adopted for NLFFF extrapolations is called preprocessing. Here we use the preprocessing code developed by \citet{2014Jiang} to get the bottom BC for the CESE-MHD-NLFFF extrapolation code before the computation is carried out.

\subsection{Grid Construction and Modified CESE-MHD-NLFFF Code for Embedded Magnetograms} \label{modified_cese}
Considering the speed and accuracy of the computation for the realistic solar magnetograms, a nonuniform grid structure within a block-structured (one block contains a group of cells) parallel computation framework has been adopted for the CESE-MHD-NLFFF code with the help of the PARAMESH software package \citep{2000MacNeice}. For the grid initialization of CESE-MHD-NLFFF code, the whole computational domain includes the pre-set central core region and the surrounding buffer region to reduce the influence of the side boundaries \citep{2013Jiang}. Then the whole computational domain is divided into blocks with different spatial resolutions, and all blocks have identical logical structures. The blocks are evenly distributed among processors. The block structures can be refined or de-refined, which provides the flexibility for embedding nonuniform magnetograms as bottom BCs. 

\begin{figure}[htb]
\centering
\includegraphics[width=0.7\textwidth]{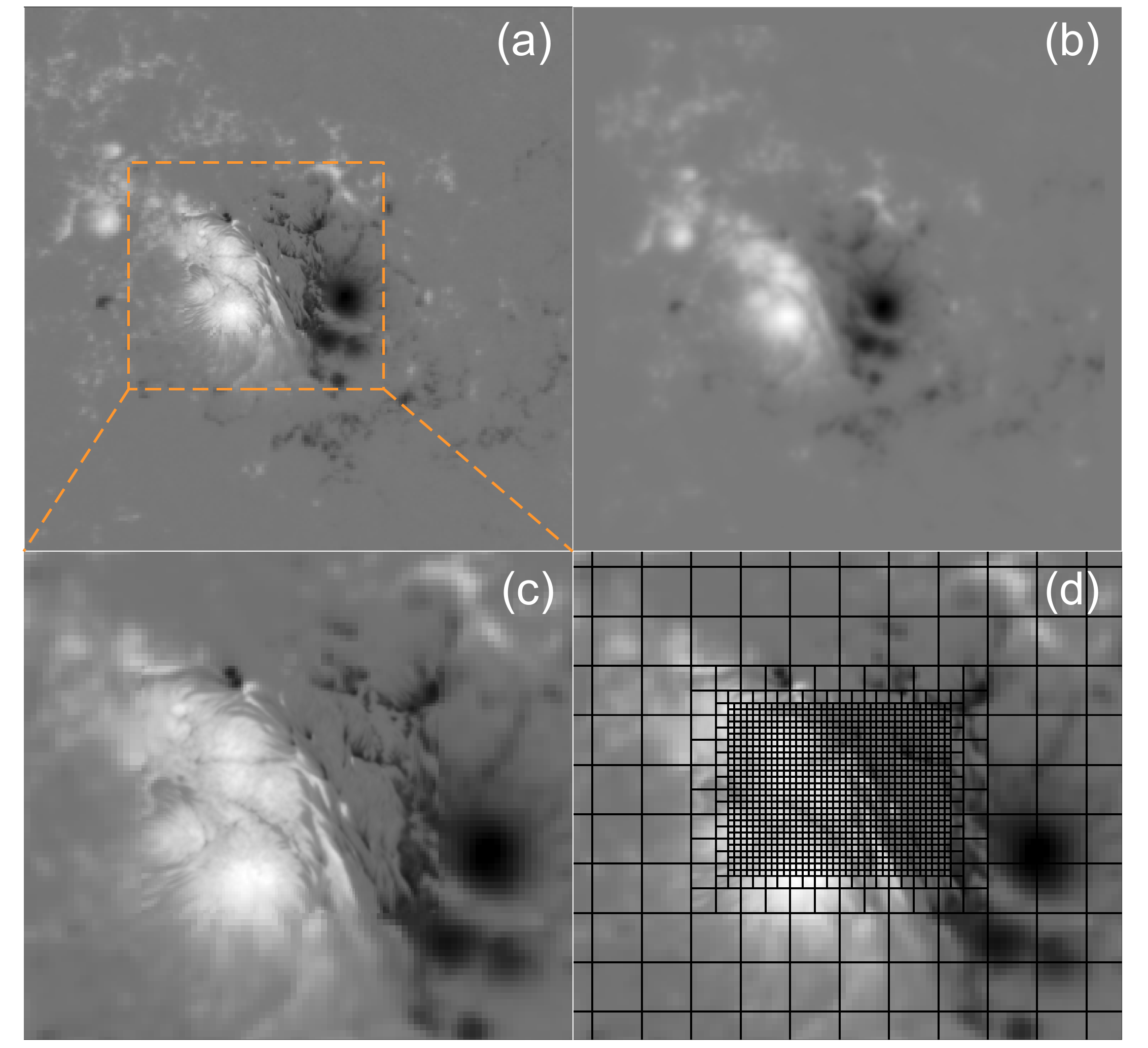}
\caption{Bottom boundary layers of $B_{z}$ component for (a) the embedded magnetogram, (b) the uniform  HMI magnetogram, and (c) a zoomed-in portion of (a) as outlined by the orange box. Panel (d) shows the associated nonuniform grid structure on the bottom boundary for (c). The whole domain is divided into  blocks with equal sides as illustrated in (d) for the bottom boundary by the solid lines, and each block  contains $8\times8\times8$ cells. The side length of the cell for  the innermost block is $0.125''$ corresponding to the side length $1''$ of the  blocks which form the core region as shown. It doubles three times to reach the cell size $1''$ for the outermost region which is the buffer region. In-between is the intermediate region.} \label{bc_grid}
\end{figure}

To apply the embedded magnetograms with nonuniform spatial resolutions for the NLFFF extrapolation, we develop a modified version of the CESE-MHD-NLFFF code to utilize the embedded map as the bottom BC. Figure \ref{bc_grid}(a) and (b) shows the difference of the bottom boundary layers between the nonuniform embedded and uniform magnetograms. Specifically, we redesign the grid structure for the whole computational domain and embed the higher-resolution magnetogram within a rectangular region into the bottom boundary, forming the core region. An enlarged version of the bottom boundary surrounding the core region is presented in Figure \ref{bc_grid}(c). Firstly, the initial grid structure should be built according to the uneven spatial resolutions of the embedded magnetogram before the computation. In PARAMESH, there is a routine to check the difference in the refinement level between the refined block and its neighboring blocks which should be no more than one level. For example, in a uniform grid structure with grid size dx = 8, the grid size in the core region cannot be refined once to dx = 1 instantly, but it can only be refined once to dx = 4. To reach the finest grid size dx = 1, two additional intermediate regions are required with the grid sizes, dx = 2 and dx = 4 (see an illustration of the intermediate regions in Figure \ref{bc_grid}(d)), respectively. Therefore, for a nonuniform embedded magnetogram (the resolution difference should be integral powers of two), we need additional intermediate regions between the central core region and the buffer region due to the constraint from the PARAMESH package. So different from the grid structure in a uniform magnetogram which mainly consists of a core region and the surrounding buffer region, the bottom boundary for an embedded magnetogram will mainly be divided into three parts: the inner core region for the higher-resolution magnetogram (in a smaller FOV), the intermediate regions from the rebinned higher-resolution magnetogram, and the surrounding buffer region populated by the lower-resolution magnetogram (with a larger FOV). One important principle for our embedding is to keep the higher-resolution magnetogram in its entirety as much as possible, so the intermediate regions between the core region and the buffer region are generally kept as narrow as possible. Once the relative positions for the two aligned magnetograms are obtained, the grid structure for the whole computational domain can be set up.

After the grid initialization, the initial solutions for all blocks in the whole computation domain will be assigned by a potential field solution derived from the higher-resolution magnetogram in the core region via the Green's function method \citep{1977Chiu}. On the bottom boundary, values from the higher-resolution magnetogram will be assigned to the innermost core region and the intermediate regions with proper rebinning. In contrast, the buffer region will adopt values from the lower-resolution magnetogram. The bottom BC is usually applied gradually, reaching the assigned values during the initial iteration steps, and then it will be fixed during the remainder of the computation.

\subsection{Convergence Study and NLFFF Quality Metrics}
To verify the quality and accuracy of NLFFF extrapolation results, routine check and evaluation of the extrapolated coronal magnetic field in a volume, including the force-freeness and divergence-freeness metrics, and comparison with coronal observations are usually required according to various validation studies of NLFFF modeling results \citep{2006Schrijver, 2008Metcalf, 2015DeRosa}. In this study, we calculate several NLFFF quality metrics to examine the trend of the extrapolation results along the relaxation process. The quality metrics include the residual of the field between two successive iteration steps, the usual force-freeness parameter \textit{CWsin}, the divergence-freeness parameter $\langle|f_{i}|\rangle$ and the total magnetic energy $E_{tot}$ \citep[see a more complete description in][] {2022He}. 

\begin{figure*}[htb]
\centering
\includegraphics[width=0.9\textwidth]{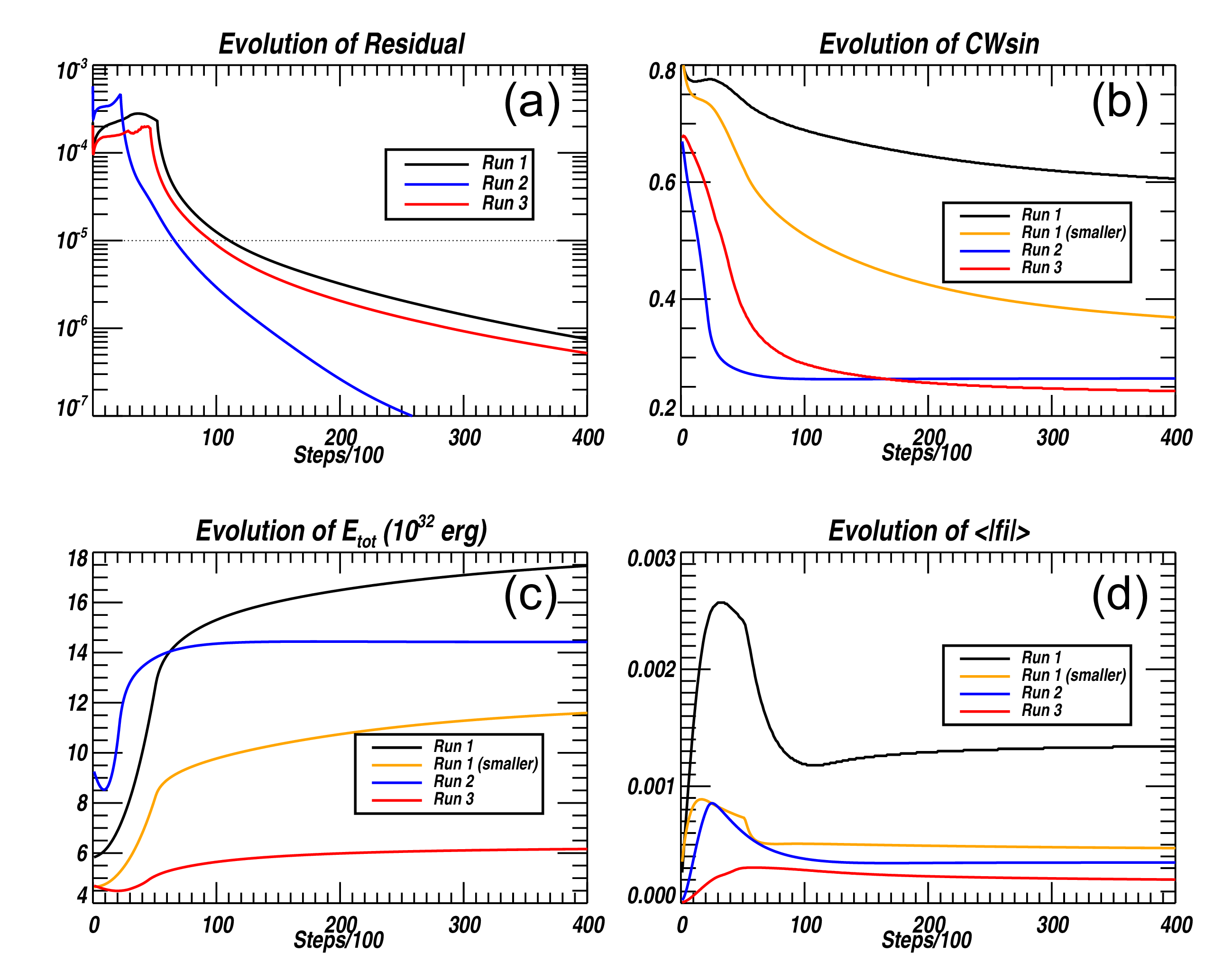}
\caption{Evolution of the convergence metrics for the three extrapolation runs: (a) the residual of the magnetic field, (b) the force-freeness paramter \textit{CWsin}, (c) the total magnetic energy $E_{tot}$, and (d) the divergence-freeness parameter $\langle|f_{i}|\rangle$. In practice, these metrics in (b)-(d) are calculated for different domains. The base areas of the domains on the bottom boundary are $192'' \times 192''$ for \textit{Run 1} and \textit{Run 2}, and $62.5''\times 53.75''$ for \textit{Run 3}, respectively. Another smaller domain with a base area of $62.5''\times 118.75''$, the same as the FOV of Figures~\ref{p1field line} and \ref{p2field line}, is applied additionally for \textit{Run 1}. All metrics are calculated with a domain height z = 300 pixels (the pixel size changes for different runs).} \label{nlfffmetrics}
\end{figure*}

For the 2015 June 22 event, we perform three extrapolation runs with different bottom BC inputs. Associated descriptions about the bottom BC inputs and grid structures are listed in Table \ref{runs_info}. The extrapolation \textit{Run 1} employs the nonuniform embedded magnetogram with a FOV of $204'' \times 204''$. The higher-resolution GST/NIRIS magnetogram with a FOV of $50'' \times 46''$ has been embedded into a larger SDO/HMI map with $1''$/pixel resolution. Two additional runs are also conducted for comparison. \textit{Run 2} is carried out based on the uniform SDO/HMI vector magnetogram in a FOV of $204'' \times 204''$, with a rebinned spatial resolution at 1$''$/pixel. \textit{Run 3} employs the uniform higher-resolution vector magnetogram from GST/NIRIS in a smaller FOV of $50'' \times 46''$ with a spatial resolution at $0.125''$/pixel. The corresponding NLFFF quality metrics are calculated in a larger domain ($192'' \times 192''$ in area, close to the size of the HMI magnetogram, for \textit{Run 1} and \textit{Run 2}) and in a smaller domain ($62.5'' \times 53.75''$, close to the size of the NIRIS magnetogram, for \textit{Run 3} only). From Figure \ref{nlfffmetrics}, the residuals from three runs all decrease to a small magnitude of the order $10^{-6}$ toward the end of the iteration. As for divergence-freeness, all three runs become stable after $\sim$ 20,000 steps. Figure \ref{nlfffmetrics}(b) shows that the convergence of the force-freeness parameter in \textit{Run 1} is more complex than the other runs. In a larger domain of $192'' \times 192''$, the \textit{CWsin} value for \textit{Run 1} keeps decreasing but it is relatively higher ($\sim$ 0.6) than the result from \textit{Run 2} within 40,000 steps. Considering the nonuniform BC we applied for \textit{Run 1}, such high \textit{CWsin} value in a large domain may be due to the difference between the part of the updated outer bottom boundary from the HMI magnetogram and the potential field solution based on the inner GST magnetogram in \textit{Run 1}. Therefore, we also calculate the \textit{CWsin} values in a smaller domain with a size of $62.5'' \times 118.75''$ on the bottom boundary (orange curves in Figure \ref{nlfffmetrics}) for consistency check, which reduces to $\sim$ 0.37 after $\sim$ 40,000 steps. This corresponds to the main part of the volume in which the subsequent topological analysis will be performed (see Section \ref{sec_results}). For reference, the \textit{CWsin} value for \textit{Run 3} becomes almost stable and equal to \textit{Run 2} after 25,000 steps. For the magnetic energy $E_{tot}$ in a larger domain, \textit{Run 2} becomes stabilized while \textit{Run 1} shows a slowly increasing trend. For a smaller domain, the evolution of $E_{tot}$ for \textit{Run 3} and \textit{Run 1} (smaller) become stabilized with a similar trend. The computation time for three extrapolation runs varies, and is generally proportional to the count of blocks as shown in Table \ref{runs_info}. The special nonuniform grid structure for \textit{Run 1} results in an intermediate run time and may also contribute to the behaviors of the quality metrics as described above.

To get a converged and consistent result for later analysis, we further check the quality of these extrapolation results with additional quality metrics for force-freeness and divergence-freeness. Here we show the comparison of NLFFF quality metrics at 40,000 iteration steps from three extrapolation runs in Table \ref{quality_metrics}. For \textit{Run 1}, the additional metrics in a smaller volume ($62.5'' \times 118.75'' \times 37.5''$) are derived. Similarly, the additional metrics for \textit{Run 2} and \textit{Run 3} are also calculated, with volume sizes of $192'' \times 192'' \times 300''$ and $62.5'' \times 53.75'' \times 37.5''$, respectively. The \textit{CWsin} values for all three runs are around 0.2-0.4, which are consistent with other extrapolation results for realistic solar magnetograms \citep{2013Jiang, 2009DeRosa}. Given that the small-scale structures in the magnetograms with weak magnetic field may increase the \textit{CWsin} value due to weak currents, we also evaluate the force-freeness and divergence-freeness with two additional metrics, $E_{\nabla \times \textbf{B}}$ and $E_{\nabla \cdot \textbf{B}}$, which estimate the residual force in the extrapolation results. The residual force comes from two parts: one is the Lorentz force $(\nabla \times \textbf{B}) \times \textbf{B}$, and the other one is the non-vanishing divergence of the magnetic field $\textbf{B}\nabla\cdot \textbf{B}$ from the numerical error \citep[see the detailed descriptions in][]{2017Duan}. The results of the two additional metrics for all extrapolation runs are small and of the same orders of magnitude and consistent with the previous reports \citep{2013Jiang, 2017Duan, 2022He}. Thus these extrapolation results extracted for the aforementioned smaller volume can be considered as converged solutions and are to be further analyzed and compared with additional observations.

\section{Results from extrapolations and observations} \label{sec_results}
\begin{figure}[htb]
\plotone{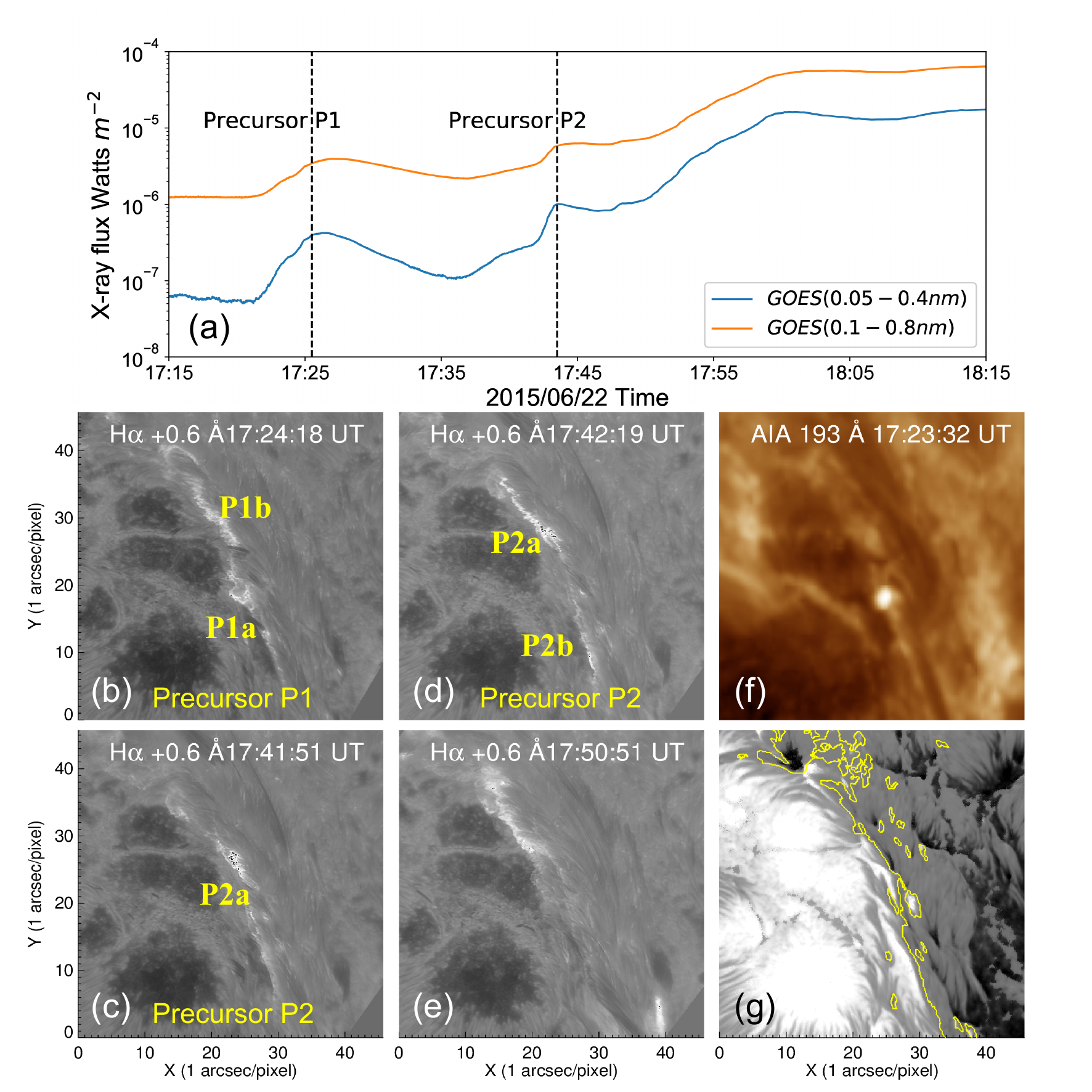}
\caption{Precursor brightenings and associated observations. Panel (a) shows the GOES X-ray flux during the precursors. Two dashed lines mark the emission times for the two precursors which occurred at $\sim$ 17:25 UT and $\sim$ 17:43 UT, respectively. (b-g) Structural evolution of the H$\alpha$ brightenings P1a/P1b and P2a/P2b from the GST observations as identified by \citet{2017Wang} before the peak of the main flare (reproduced from Figure 1 in \citet{2017Wang}), and (f) the corresponding image observed in the SDO/AIA 193 \r{A} wavelength. Panel (g) shows the corresponding GST/NIRIS magnetogram of $B_{z}$ component. Yellow contour marks the polarity inversion line (PIL).} \label{fp1p2}
\end{figure}

Before the main flare eruption, the two small-scale precursor brightenings were identified as P1 and P2 near the PIL at $\sim$ 17:24 UT (P1) and $\sim$ 17:42 UT (P2) from the study utilizing the high-resolution GST observations by \citet{2017Wang}. Part of the results from that analysis is reproduced in Figure \ref{fp1p2}(b-g). Figure \ref{fp1p2}(a) shows the GOES X-ray flux during the flare precursors. There are two small peaks appearing before the main flare eruption. These impulsive emission times of the GOES X-ray flux also coincide with the corresponding H$\alpha$ brightenings as marked. Figure \ref{fp1p2}(b-g) are regenerated from \citet{2017Wang} to show the structural evolution of the flare precursors from the high-resolution GST/H$\alpha$ observations. As shown in (b) and (d), the brightening P1b and P2a (``a'' and ``b'' for each precursor period are named by \citealt{2017Wang} based on the chronological order of their occurrence times) are almost co-spatial while the P1a lies southward in a relatively different area from the brightening P2b. The corresponding GST/NIRIS magnetogram of $B_{z}$ component is presented in (g) and overplotted with the PIL (colored by the yellow contour). During the precursor P1, a brightening point was also observed in the coaligned SDO/AIA 193 \r{A} image in (f), which located close to the P1a brightening region.

\subsection{Field Line Connectivity and GST Observations} 
\begin{figure*}
\includegraphics[width=\textwidth]{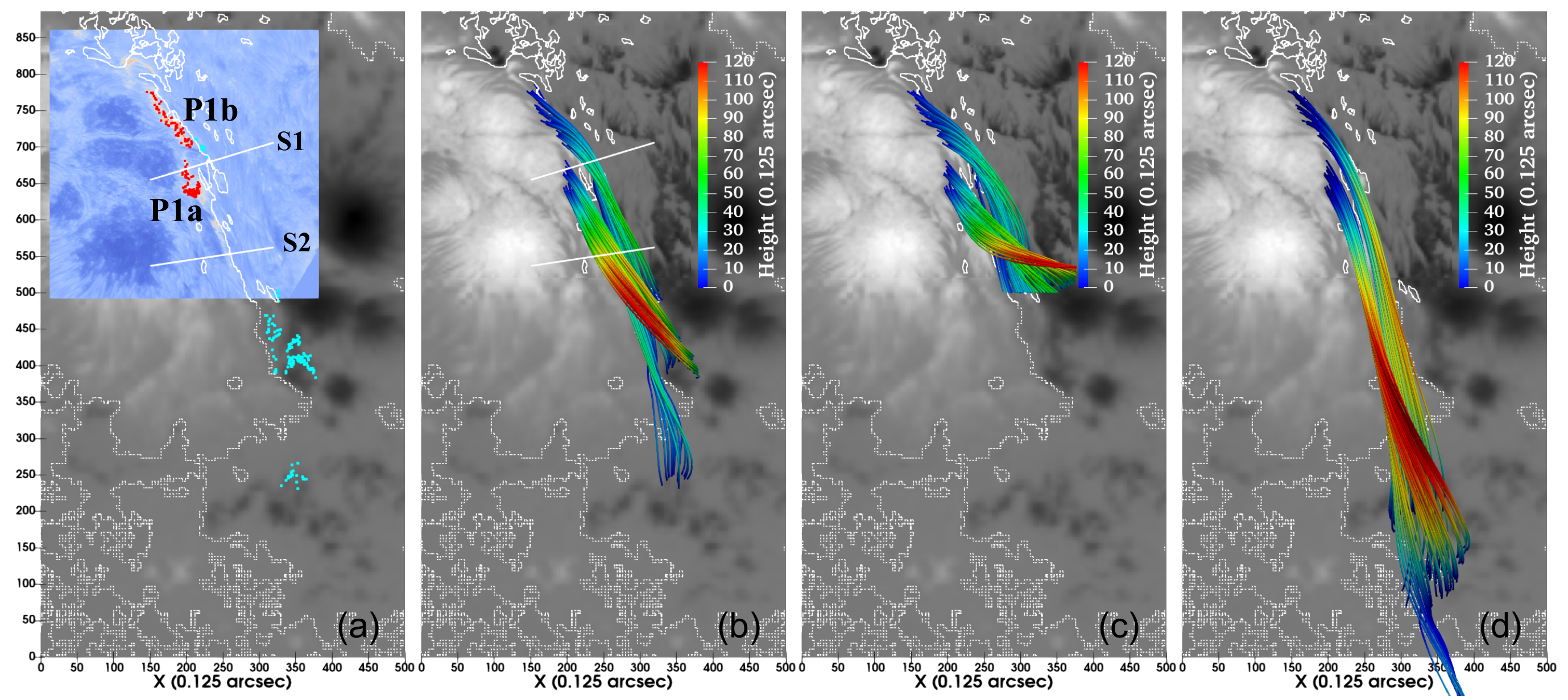}
\caption{Analysis of selected magnetic field line connectivity near the precursor brightening P1. (a) Two groups of brightening pixels from P1a and P1b are selected over the GST/H$\alpha$+0.6\r{A} image (in light blue shades) at 17:24:18 UT together with their conjugate field-line footpoints based on the extrapolation result from \textit{Run 1}. They are color-coded by the  magnetic polarity of the corresponding field-line footpoints on the bottom boundary: positive in red and negative in cyan. Panels (b-d) show the field lines that originate from the red points in (a) for all three NLFFF extrapolation runs (Run 1, 3, and 2, respectively) which are colored by the height. In each panel, the $B_{z}$ map for \textit{Run 1} on the bottom boundary is drawn in gray scales with the PIL indicated by the white contours. In (a) and (b), the two white lines S1 and S2 indicate the positions of two vertical slices to be displayed in Figures \ref{twist_s1} and \ref{twist_s2}.} \label{p1field line}
\end{figure*}

To have a better understanding of the fine-scale structures in the 3D volume for the precursor brightenings, we compare the available GST/H$\alpha$ observations with the 3D coronal magnetic field topology from the static extrapolations \textit{Run 1} to \textit{Run 3}, respectively. Figure \ref{p1field line} illustrates the selected magnetic field line connectivity near precursor P1. Firstly, two areas of interest for regions P1a and P1b are identified from the GST/H$\alpha$ image at 17:24:18 UT in light blue scales (the same as Figure \ref{fp1p2}(b)) when the brightening intensity of a pixel is greater than a certain threshold in Figure \ref{fp1p2}(b). Two groups of such brightening pixels from P1a and P1b are selected and colored by red (cyan) corresponding to the positive (negative) magnetic field polarity. The majority of the pixels are in red with positive magnetic polarity, and the corresponding conjugate footpoints in cyan are marked across the PIL to the south based on the \textit{Run 1} result. Then the field lines that originate from the red points for all three NLFFF extrapolation runs are drawn for comparison, and are colored by the vertical height along each field line in (b)--(d).

Most field lines from \textit{Run 3} in (c) are short sheared arcades appearing within the FOV of the GST observation. And the conjugate footpoints for the field lines originating from P1a mainly attach to a major negative polarity region of the background GST magnetogram, while field lines from P1b are generally open exiting the limited computational domain. The field lines from \textit{Run 2} are less sheared and extend longer than those results from the other two runs. Those long field lines in (d) come from P1a and P1b extending to an area beyond the FOV of the GST magnetogram and further southward of the cyan dots in (a). It is shown that some field lines from P1a in (d) are not closed within the selected domain as shown. In contrast, the corresponding field lines from \textit{Run 1} in (b) show conjugate negative polarity footpoints (cyan dots) extending beyond the FOV of the GST magnetogram, but well within the FOV of the HMI magnetogram and across the main PIL between the two major positive and negative polarities. The group of selected field lines from P1b in (b) stays closed on the bottom boundary while another group of field lines from P1a appears as a shorter sheared arcade and lies above the group that originates from P1b. We can see that the field line bundles in all runs reach a similar maximum height of $\sim 15''$. From these comparisons, \textit{Run 1} shows a reasonable consistency for the sheared arcade structures shown across the main PIL where the two strong polarity regions are separated based on the GST and HMI magnetograms. The results for the other two runs are clearly affected by the sizes of their computational domains with each maintaining a uniform grid setting.

\begin{figure*}
\includegraphics[width=\textwidth]{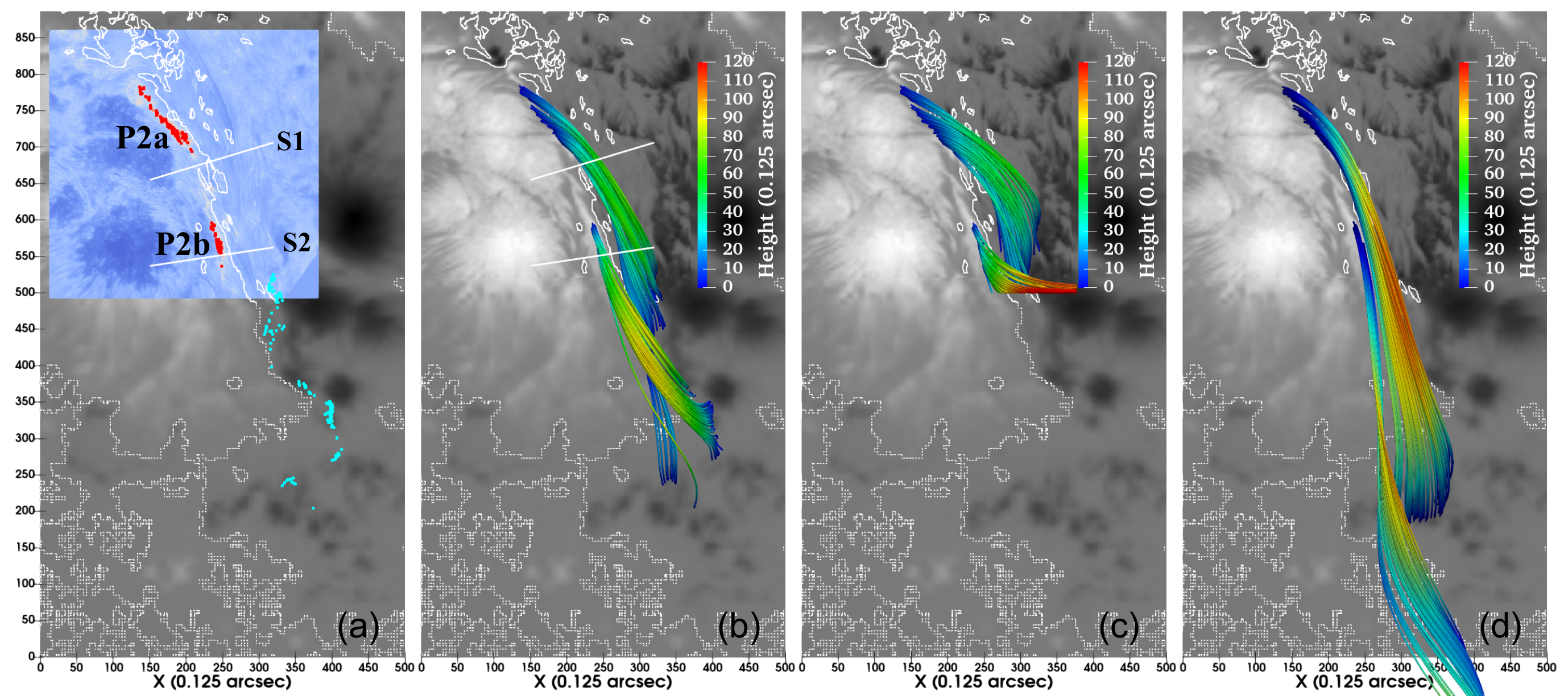}
\caption{Similar analysis for the selected field lines corresponding to the precursor brightening P2, based on the GST/H$\alpha$+0.6\r{A} image at 17:42:19 UT and the extrapolation result from \textit{Run 1}. Format is the same as Figure \ref{p1field line}.} \label{p2field line}
\end{figure*}

For precursor P2, we also draw the field lines originating from chosen brightening pixels based on the similar criteria for the GST/H$\alpha$ observation at 17:42:19 UT, as presented in Figure \ref{p2field line}. Similar to Figure \ref{p1field line}, the field lines from \textit{Run 1} in Figure \ref{p2field line}(b) show the magnetic structures corresponding to the precursor brightenings with both a fine scale and a spatial extent along the main PIL confined within the strong field regions albeit at lower heights. The H$\alpha$ observations show that the brightening region P2a is nearly co-spatial with P1b, whereas another region P2b lies in a different area from P1a. In terms of the height distribution, the field line bundle originating from P2b situates in a lower height comparing to the field lines from P1a while the field lines from P1b and P2a have similar heights. As a general feature, the two groups of field lines from \textit{Run 1} lie almost parallel to the PIL. What's more, the positive polarity footpoints of the sheared field line bundle from P2b is close to a part of the conjugate negative footpoints from P2a, as indicated by the red and cyan points in Figure \ref{p2field line}(a), which is a potential configuration favorable for magnetic reconnection. Further topological analysis result will be presented in Section \ref{topology}.

\subsection{Field Line Connectivity and AIA Observations}
\begin{figure}[htb]
\epsscale{1.1}
\plotone{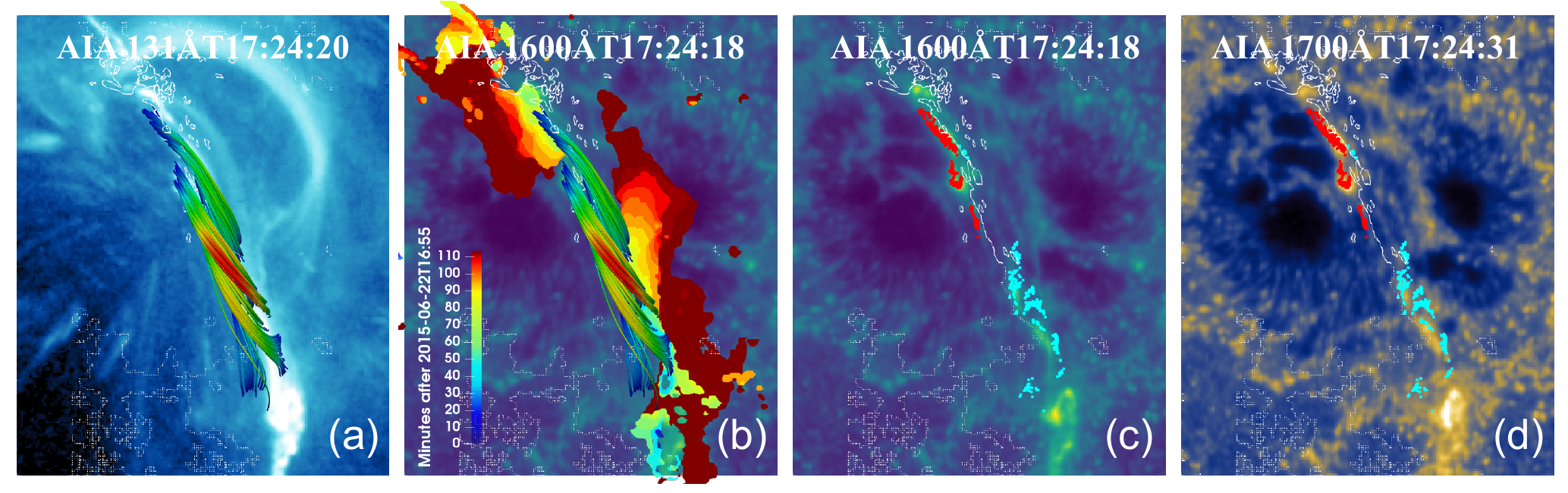}
\caption{Comparison between the field line connectivity and AIA observations in different wavelengths. (a) All selected field lines from \textit{Run 1} in Figures \ref{p1field line} and \ref{p2field line} are superimposed on the AIA 131 \r{A} observation. (b) Contours of flare ribbons colored by elapsed time since 16:55 UT (see the color bar) are overplotted with the set of field lines over an AIA 1600 \r{A} image. (c-d) The footpoints with positive (red) and negative (cyan) magnetic polarity  for the set of field lines are drawn over the corresponding  AIA 1600 and 1700  \r{A} images.} \label{aia_comp}
\end{figure}

Due to the constraint of limited FOV, the fine-scale GST observation is not available in a larger FOV encompassing the pairs of conjugate field line footpoints in the negative polarity regions. To further verify the identified conjugate negative footpoints from the extrapolation \textit{Run 1}, we compare the field line connectivity with the corresponding \textit{SDO}/AIA observations in a larger FOV to find its connection to the precursor brightenings. In Figure \ref{aia_comp}, a series of AIA observations during the precursor P1 are used as the background images along with the extrapolated field line bundles from \textit{Run 1}. In (a), the field lines which originate from the selected H$\alpha$ brightening pixels during precursor P1 and P2 are overplotted and they have a good consistency with the central hot loops in AIA 131 \r{A} at 17:24 UT. In addition, in (b), the flare ribbon brightenings are also marked and color-coded by elapsed time since 16:55 UT. The positive and negative footpoints of identified field lines from precursors P1 and P2 are overlaid to AIA observations in (c)-(d). During precursor P1, the positive footpoints (red) near P1a are co-spatial with the brightening patches observed simultaneously in AIA 1600 \r{A} and 1700 \r{A} wavelengths. And the identified negative footpoints from \textit{Run 1} overlap partially with the brightening patches to the south in (c) and (d). Furthermore, as the flare ribbons can be used as an estimation for footpoints of reconnected magnetic field lines \citep{2002Qiu, 2004Qiu}, the overall flare ribbon evolution is superimposed in Figure \ref{aia_comp}(b). It indicates the initiation of the main flare reconnection closer to the PIL at the earlier times and the subsequent extension of the ribbons away from the PIL when reconnection proceeds during the flare main phase. The multiple bundles of field lines identified from precursors P1 and P2 have most footpoints locate inside or near the flare ribbons at earlier times, which offers additional evidence in support of this magnetic field configuration for the precursor magnetic reconnection followed by the main phase flare reconnection \citep[see, e.g.,][]{2001Moore}.

\subsection{Additional Topological Analysis} \label{topology}
From the previous comparisons between the extrapolations and observations, several groups of sheared arcades over the main PIL have been successfully reconstructed corresponding to the H$\alpha$ precursor brightenings. And the conjugate footpoints of precursor brightenings based on the extrapolation \textit{Run 1} also show consistency with the alternative and subsequent brightening regions in the AIA observations. But it remains a question: how to find the potential sites for magnetic reconnection? How do precursor brightenings evolve and what is the subsequent reconnection sequence? To look into such questions, especially the first one, additional parameters like the normalized current density ($|\mathbf{J}|/|\mathbf{B}|$), the magnetic twist number $T_{w}$, and the squashing degree Q are calculated to analyze the magnetic topology in a specific volume. The magnetic twist number $T_{w}$ gives a good estimation of how many turns two infinitesimally close field lines wind about each other \citep{2006Berger, 2016LiuR}. And the squashing degree Q quantifies the change of magnetic connectivities \citep{1996Demoulin, 2002Titov}. For example, complex 3D magnetic structures can be distinguished near high Q regions, where the gradient of the field line connectivity as measured by the Q value is large. 

\begin{figure}[htb]
\epsscale{0.8}
\plotone{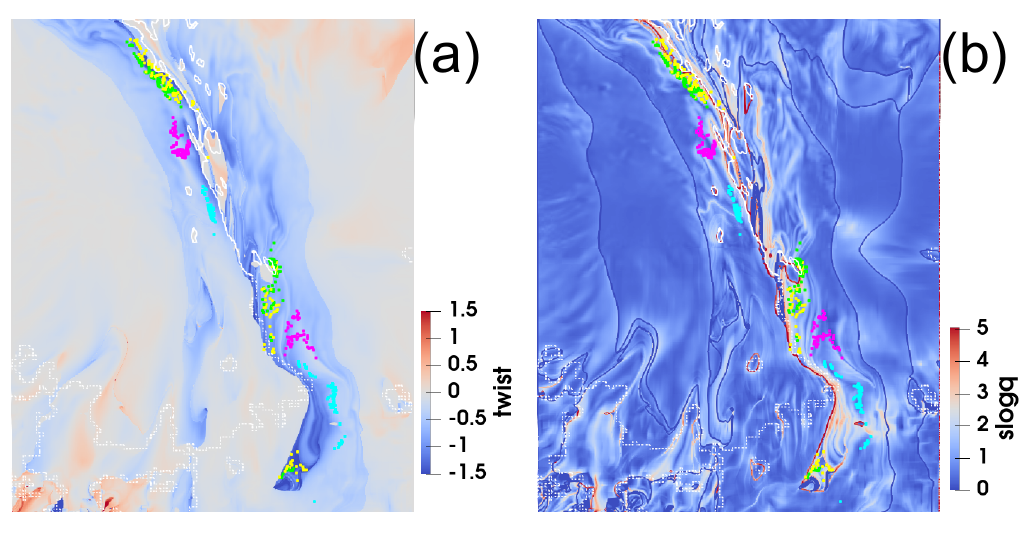}
\caption{The top views of the distributions of (a) the twist number $T_{w}$  and (b) the squashing degree Q (in base-10 logarithmic scale) on the bottom boundary. The footpoints for the four groups of field lines illustrated in   Figures \ref{p1field line} and \ref{p2field line} are overlaid in different colors: precursor P1b in yellow, P1a in  purple, P2a in green, and P2b in cyan.} \label{tw_logq}
\end{figure}

\begin{figure}[htb]
\epsscale{0.8}
\plotone{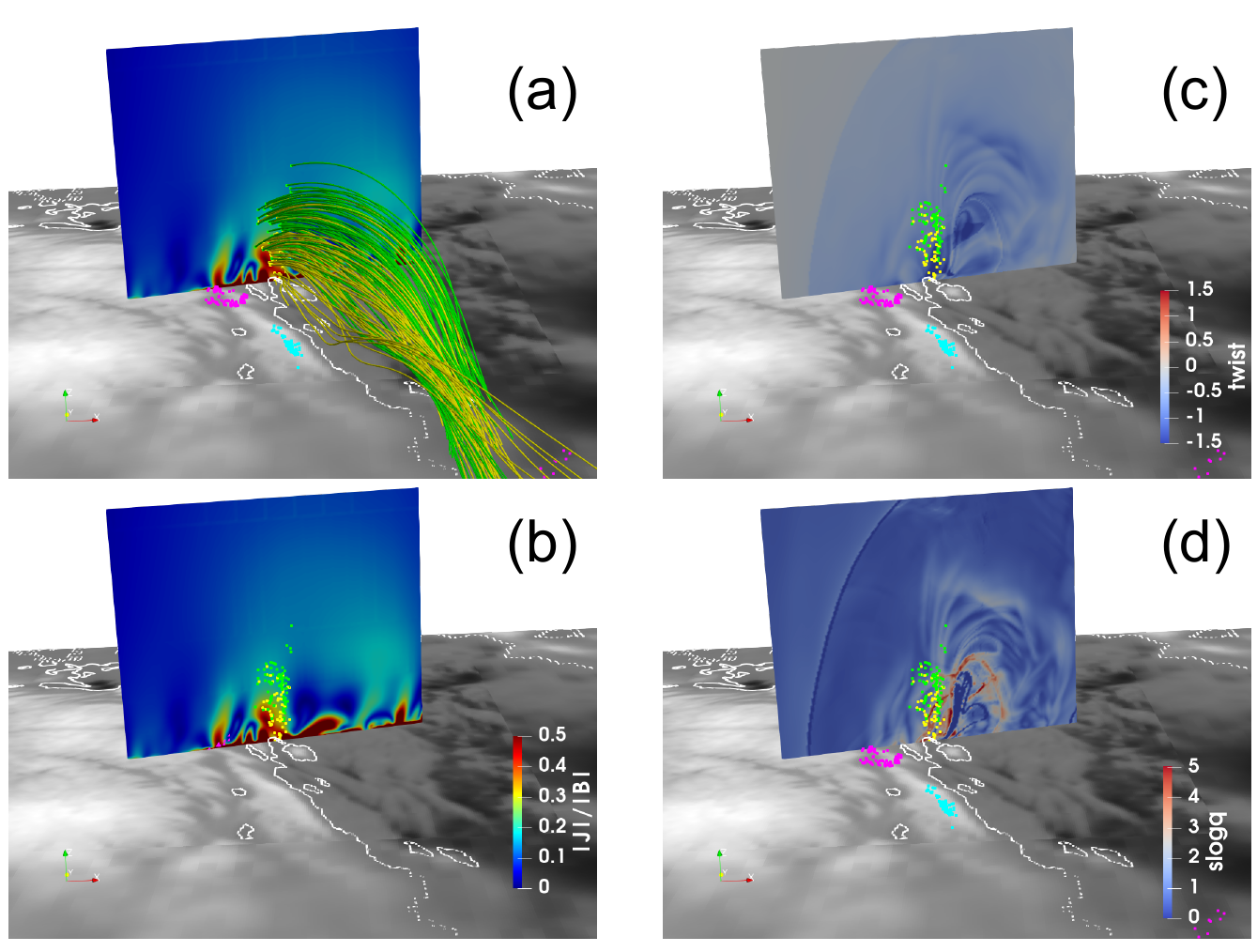}
\caption{ Topological analysis of the field lines corresponding to brightenings P1 and P2. (a-b) The distribution of $|\mathbf{J}|/|\mathbf{B}|$ on the vertical slice S1 with the interception points for the identified field lines from precursor P1b (yellow), P1a (purple), P2a (green) and P2b (cyan). (c-d) The distribution of the twist number $T_{w}$ and the squashing degree Q on the vertical slice S1 with overplotted field line interception points. The corresponding $B_{z}$ map with the PIL highlighted by white contours is drawn on the bottom layer.} \label{twist_s1}
\end{figure}

\begin{figure}[htb]
\epsscale{0.8}
\plotone{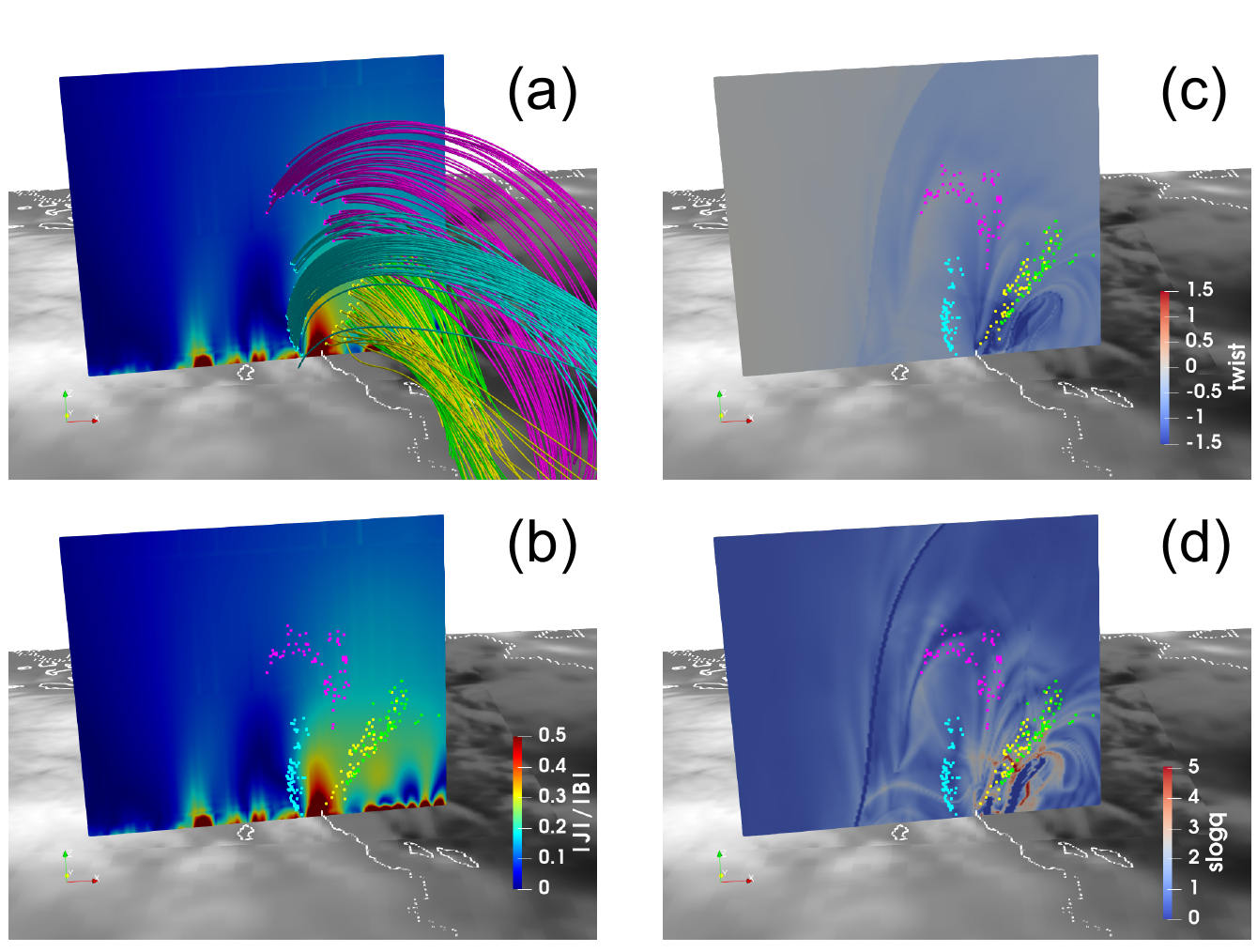}
\caption{Similar analysis as shown in Figure \ref{twist_s1}  for the  distributions of the corresponding toplogical parameters on the vertical slice S2. Format is the same as Figure \ref{twist_s1}. } \label{twist_s2}
\end{figure}

The top views of the twist number $T_{w}$ and the squashing degree Q distributions on the bottom boundary are shown in Figure \ref{tw_logq} along with the footpoints from the identified field lines corresponding to the precursor brightenings P1 and P2. They show again the spatial distribution of the field-line footpoints for P1 and P2 along and within the extent of the main PIL which is characterized by a sharp ``ridge'' like feature in these parameters. Figures \ref{twist_s1} and \ref{twist_s2} show the distributions of $|\mathbf{J}|/|\mathbf{B}|$, the twist number $T_{w}$ and the squashing degree Q on the two vertical slices, S1 and S2, as marked in Figures \ref{p1field line} and \ref{p2field line}. The extrapolation result based on \textit{Run 1} is obtained with the embedded vector magnetograms at 17:32 UT (GST) and 17:36 UT (HMI), which provides a snapshot of the magnetic field topology at a time between precursors P1 and P2. For the slice S1 in Figure \ref{twist_s1}, the field lines from P1b and P2a (colored by yellow and green respectively) go through a region with relatively high current density. Besides, the squashing degree Q around those field lines exhibits a complex pattern intermixed with high values. Such complexity also remains around the intercepting field-line points on the slice S2 in Figure \ref{twist_s2}(d), indicating the potential sites for magnetic reconnection between these field lines, which could result in the co-spatial brightenings at their footpoints as observed in P1b and P2a areas. Considering the magnetic topology near other brightenings (P1a and P2b), the field line bundles originating from P1a and P2b (pink and cyan) are next to the other two with modest Q values in Figure \ref{twist_s2}(d). The twist numbers are all insignificant for these field lines. However, from the distribution of the intercepting points of different field lines on the slice S2 in Figure \ref{twist_s2}(b), the cyan field line bundle from P2b is lower than the pink bundle from P1a and it is closer to yellow/green field line bundles from P1b/P2a. In addition, the cyan field lines from P2b and green field lines from P2a are separated by a high $|\mathbf{J}|/|\mathbf{B}|$ region in Figure \ref{twist_s2}(b). Such a configuration may correspond to the initial stage of the tether-cutting reconnection scenario described in \citet{1989vanBallegooijen, 2001Moore}. That is the magnetic reconnection among and between the sheared magnetic flux bundles as shown across the main PIL may take place, resulting in the brightenings of the associated field line footpoints without significant changes of their positions. And this could be one result to explain the spatial features consistent with the observed brightenings, but not the temporal changes. The purpose of an extrapolation is to provide a snapshot at a specific time. To examine the temporal change in the magnetic field topology is beyond its capability.
Nonetheless, we also performed an additional extrapolation run for the embedded magnetogram around 17:48 UT, at a time that is a few minutes after P2. The result shows similar magnetic field line topology for the precursor regions as we have presented in this section. It probably implies that the reconnection associated with the precursors only involved small amounts of flux and the reconnection did not significantly change the flux distributions of the precursor regions. 

\section{Summary and Conclusions} \label{sec_summary}
In this study, we have applied the CESE-MHD-NLFFF extrapolation method to a nonuniform embedded magnetogram for the first time to study the fine-scale structures of  precursors before the main flare eruption.
Three extrapolation results are obtained with different bottom BCs and grid structures, namely, \textit{Run 1} with a nonuniform embedded magnetogram, \textit{Run 2} with a uniform SDO/HMI magnetogram, and \textit{Run 3} with a uniform GST/NIRIS magnetogram. In the convergence study, the residual and divergence-freeness parameter for all three runs become sufficiently small during the iteration, while the force-freeness parameter shows more complicated behaviors. The \textit{CWsin} value in a larger computational domain for \textit{Run 1} is higher than the other results, but it reduces to $\sim$ 0.37 for a smaller volume in which the magnetic field topology is examined in detail. The deviation from a strict force-free state in \textit{Run 1} could be due to the nonuniform BC and grid structure designed for the embedded magnetogram. Nonetheless, the \textit{CWsin} values for all three runs after $\sim$ 40,000 iteration steps are around $0.2-0.4$, which are consistent with prior extrapolation results that are considered to be converged solutions for realistic solar magnetograms. After the converged results are obtained, we look into the reconstructed 3D magnetic field topology around the precursor brightenings, and compare the field line connectivity with the GST/H$\alpha$ and SDO/AIA observations. Additional topological features for the extrapolation \textit{Run 1} are investigated by focusing on the fine-scale structures around the precursor brightenings and across the main PIL. The main results are listed as follows:

\begin{enumerate}

\item For all three extrapolation runs, the field line connectivity around the precursor brightenings is compared with the GST/H$\alpha$ observations. The magnetic field lines originating from the precursor brightening regions based on \textit{Run 1} exhibit a configuration of the fine-scale magnetic structures beyond the small FOV of GST but confined within the larger FOV of HMI, more consistent with the spatial extent of the main PIL between two major magnetic polarities. Multiple sheared flux bundles are found overlying across the main PIL with groups of footpoints rooted in the positive magetic polarity regions and coinciding with each set of the observed H$\alpha$ brightening patches, P1 and P2, respectively. 

\item The selected field line bundles originating from the H$\alpha$ brightening patches from \textit{Run 1} show an overall shape consistent with the corresponding AIA observations in different wavelengths. Those selected field lines have a good correspondence with the hot loops observed in AIA 131 \r{A} passband. And their footpoints are attached to the inner sides of the flare ribbons with the closest distances from the PIL, which indicates a potential configuration for the magnetic reconnection during the flare precursors at earlier times. 

\item With the magnetic field topological analysis near the precursor brightenings based on the extrapolation \textit{Run 1}, including the distributions of the normalized current density $|\mathbf{J}|/|\mathbf{B}|$, the magnetic twist number $T_{w}$ and the squashing degree Q, a plausible configuration for magnetic reconnection is found. Such structures may correspond to the initial stage of the tether-cutting reconnection scenario, before the main ``flare onset'', for instance. 

\end{enumerate}

These results based on \textit{Run 1} represent the application of the CESE-MHD-NLFFF extrapolation method for an embedded photospheric magnetogram from the GST/NIRIS and SDO/HMI observations. By utilizing different analyzing tools for the extrapolation results together with additional observations, the fine-scale magnetic structure around flare precursors are found to be consistent with the associated high-resolution GST/H$\alpha$ observations. We conclude that the reconstructed magnetic field line topology/connectivity across the main PIL from \textit{Run 1} is more plausible for the subsequent magnetic reconnection among the sheared flux bundles, resulting in the corresponding precursor brightenings.  We thus provide a viable approach to investigate the fine-scale structures associated with solar eruptions  by combining the  high-resolution magnetogram in a smaller FOV with another set of magnetogram in a larger FOV. By resolving the potential site for the small-scale precursors before the main flare eruption, this study demonstrates the merit of employing the ultra high-resolution magnetogram with its native resolution. The reconstructed magnetic field over the whole computation volume could also be further analyzed and the results could contribute to improving our understanding on how to make a connection between the small-scale energy release processes and the main phase of solar eruptions at  larger scales, including the filaments, flares, CMEs and so on. This will be pursued in future studies.

\begin{acknowledgements}
We thank the teams of BBSO/GST and SDO for providing the valuable data of this event. W.H. and Q.H. acknowledge support from the National Science Foundation grants AST-2204385, AGS-2050340 and AGS-1954503. W.H. is also supported by the DKIST Ambassador Program, funding for which is provided by the National Solar Observatory, a facility of the National Science Foundation, operated under Cooperative Support Agreement number AST-1400450. J.J. and H.W. are supported by NSF awards AST-2204381 and AGS-1954737. S.S.N. acknowledges the NSF-AGS-1954503 and NASA-LWS-80NSSC21K0003 grants. We are grateful to Dr. Jiong Qiu for providing the code for the flare ribbon analysis. We gratefully acknowledge the use of data from the GST of BBSO. BBSO operation is supported by US NSF AGS-2309939 and AGS-1821294 grants and New Jersey Institute of Technology. GST operation is partly supported by the Korea Astronomy and Space Science Institute and the Seoul National University.
\end{acknowledgements}

%

\vspace{5mm}
\facilities{SDO, GST, GOES}










\bibliography{sample631}{}
\bibliographystyle{aasjournal}

 \begin{deluxetable}{cccccc}
\tablecaption{Boundary Conditions (BCs) for Different NLFFF Extrapolation Runs\label{runs_info}}
\tablewidth{0pt}
\tablehead{
\colhead{Runs} &\colhead{Bottom BCs} & \colhead{Resolution} & \colhead{FOV of the} &Count  & \colhead{Computation Time}\\
\colhead{} & & &bottom BC &of blocks& \colhead{(to  40,000 steps)}
}
\startdata
Run 1& Nonuniform& Outer region: $1 ''$& $204'' \times 204''$ &2056&19 hrs \\
& embedded magnetogram & Core region: $0.125 ''$ & $50'' \times 46''$&\\
\hline
Run 2& Uniform SDO&$1 ''$&$204'' \times 204''$&1320&12.5 hrs\\
&magnetogram at 17:36 UT&&&& \\
\hline
Run 3& Uniform GST &$0.125 ''$ &$50'' \times 46''$&5848&52.5 hrs \\
&magnetogram at 17:32 UT&&&& \\
\enddata
\tablecomments{all computations are performed with 19 cores on a 24-core local desktop with 48 GB memory.}
\end{deluxetable}

\begin{deluxetable}{ccccc}
\tablecaption{NLFFF Quality Metrics for Force-freeness and Divergence-freeness \label{quality_metrics}}
\tablewidth{0pt}
\tablehead{
\colhead{Runs} &\colhead{\textit{CWsin}} & \colhead{$\langle|f_{i}|\rangle$} & \colhead{$E_{\nabla \times \textbf{B}}$} & \colhead{$E_{\nabla \cdot \textbf{B}}$}
}
\startdata
Run 1 (smaller)&0.369&$4.71\times 10^{-4}$&0.246&$1.54 \times 10^{-2}$\\
Run 2 &0.264&$3.47\times 10^{-4}$&0.164&$1.67 \times 10^{-2}$\\
Run 3 &0.243&$2.02\times 10^{-4}$&0.159&$1.99 \times 10^{-2}$\\
\enddata
\end{deluxetable}

\newpage

\end{document}